# TRABZO: a novel combined model for heavy-ion fusion/capture accounting for zero-point shape oscillations and dissipative effects

M. V. Chushnyakova

*Physics Department, Omsk State Technical University, 644050 Omsk, Russia*

*e-mail:* maria.chushnyakova@gmail.com

The fusion/capture cross sections (CSs) of complex nuclei (heavy ions) are often described in the literature within two approaches: i) the coupled channels model accounting for structure of the colliding nuclei and ii) trajectory models with friction and thermal fluctuations (dissipative effects). The first approach does not account for friction whereas the second one is not able to deal with sub-barrier CSs. In the present work, we have developed an algorithm for calculating the capture CSs of collision of spherical nuclei accounting for both zero-point oscillations (ZPO) of the nuclear shapes (structure effects) and for dissipative effects; i.e., in a sense we have combined the above two approaches. The bare nucleus-nucleus potential is evaluated using the semi-microscopic double-folding model with M3Y-Paris nucleon-nucleon forces. The nucleon densities are taken from the IAEA data base. For each collision partner several deformations of quadrupole and octupole type are accounted for with the probabilities corresponding to the harmonic oscillator at the ground state. In these calculations, the multipole expansion of the densities with the subsequent Fourier transforms of the multipole components and of the nucleon-nucleon forces is applied. The dissipative effects enter into our combined approach within the surface friction model well known in the literature. The final fate of a trajectory is decided by means of quantum transmission coefficients. There are two fitting parameters in the model, $\tau$ and $K_R$. Parameter $\tau$ reflects to what extent ZPO survives when the reagents approach each other. Parameter $K_R$ is the friction strength for the radial motion. All calculations have been performed for $^{16}$O+$^{92}$Zr reaction. The calculated CSs and barrier distribution are found in good agreement with the precision experimental data at reasonable values of $\tau$ and $K_R$. Calculations show that the CSs are mostly sensitive to $\tau$ at low (sub-barrier) collision energies whereas the value of $K_R$ is important at the above barrier energies. We hope the model can be upgraded and applied for describing the CSs in other reactions.

## I. INTRODUCTION

Collision of two complex nuclei with the charge numbers $Z_P$ (projectile) and $Z_T$ (target) at the energies around the Coulomb barrier often results in capture of the nuclei into orbital motion. Provided $Z_P Z_T < 800$ [1] this capture follows by fusion with formation of a compound nucleus. Thus, the capture and fusion cross sections (CSs) are very close. In the literature these CSs often are evaluated within two theoretical approaches: 1) the coupled channels (CC) model accounting for structure of the reagents [2-4] and 2) the trajectory model (TM) accounting for friction and thermal fluctuations [5-7]. Both approaches possess some shortcomings. The CC model does not account for dissipation, i.e., conversion of the kinetic energy of mutual motion into the excitation energy of the reagents. Yet the CC model utilizes the incoming wave boundary condition implying extremely strong friction when the potential barrier of the interaction energy has been overcome. Moreover, the bare Strong nucleus-nucleus Potential (SnnP) in this analysis has usually the Woods–Saxon shape with the parameters adjusted to reproduce the measured above–barrier fusion cross sections [2, 8, 9]. The diffusenesses of this profile exceed significantly the ones extracted from the analysis of elastic scattering data [10].

A possible solution of the problem of the "apparently large diffuseness" had been suggested in the TM [7], and in [11] a significant array of the precision experimental fusion cross sections had been successfully described using the TM. However, the TM being a classical model is not able to deal with sub-barrier fusion. Moreover, it did not deal with the deformed nuclei.

The quantum diffusion model (QDM) [12-14] represents a sort of symbiosis of the above two models combining their advantages, i.e., accounting for dissipation, quadrupole deformations, and orientation of colliding nuclei. This model was successful in describing the fusion CSs for many reactions [14-16]. The shortcoming of the QDM model is that it utilizes the nucleon densities which do not produce correct experimental charge densities. We consider this circumstance as an intrinsic contradiction of this model. Also, only quadrupole deformations of the nuclei are accounted for whereas in [17-19] the importance of the octupole mode was demonstrated.

About forty years ago, it was shown [20-22] that accounting for the zero-point oscillations (ZPO) of the nuclear shapes substantially remedy the agreement of the evaluated capture cross-sections in heavy-ion collisions with the data. In [23, 24] the ZPO approach has been elaborated as follows: i) the SnnP is evaluated using the semi-microscopical double folding model (DFM) with the M3Y-Paris nucleon-nucleon forces; ii) the zero-point oscillations are accounted for in a more detail way and for both collision partners. It is important to use the Paris version of the M3Y NN-forces [25] in the DFM calculations. In some recent works, the Reid-M3Y forces [26] are used [27, 28.] Yet it is stated in [25] that "the Paris potential is based on a more fundamental theory of NN-scattering than the earlier potentials... The Reid soft-core potential is based on earlier and partially erroneous phase-shift data..."

The purpose of the present work is to combine the ZPO-approach with the trajectory model in order to account for simultaneously the structure of the colliding nuclei and dissipation of the collective kinetic energy.

The paper is organized as follows. Section II is devoted to the exposition of the ZPO model. The way the SnnP is calculated is discussed in Section III. The combined model TRABZO (Trajectory model with Barrier penetration and Zero-point Oscillations) is described in Sections IV. Comparison with the experimental CSs and barrier distributions for reaction $^{16}$O+$^{92}$Zr is presented in the same section. Conclusions are formulated in Section V. Some details of calculations are presented in Appendixes.

## II. OUTLINE OF THE ZPO MODEL

The basic idea of ZPO approach is to accept the model of two independent harmonic oscillators for the quadrupole and octupole vibrational modes of a spherical nucleus. Thus, accounting for ZPO makes spherical nuclei virtually to be deformed. Here we account for both quadrupole and octupole deformations of the projectile (P) and target (T) nuclei which are characterized by the deformation parameters $\beta_{2P}, \beta_{3P}, \beta_{2T}, \beta_{3T}$.

The probabilities $\Pi_{2Pi}, \Pi_{3Pj}, \Pi_{2Tl}, \Pi_{3Tn}$ for projectile and target to possess at the ground state the deformations $\beta_{2Pi}, \beta_{3Pj}, \beta_{2Tl}, \beta_{3Tn}$ read

$$\Pi_{2Pi} = N_{2P}^{-1} \exp\left(-\frac{\beta_{2Pi}^2}{2S_{2P}^2}\right), \qquad \Pi_{3Pj} = N_{3P}^{-1} \exp\left(-\frac{\beta_{3Pj}^2}{2S_{3P}^2}\right), \tag{1}$$

$$\Pi_{2Tl} = N_{2T}^{-1} \exp\left(-\frac{\beta_{2Tl}^2}{2S_{2T}^2}\right), \qquad \Pi_{3Tn} = N_{3T}^{-1} \exp\left(-\frac{\beta_{3Tn}^2}{2S_{3T}^2}\right), \tag{2}$$

The rms deviations $S_{2P(T)}$ and $S_{3P(T)}$ are tabulated in [29] where they are denoted $\beta_2^0$ and $\beta_3^0$. The normalization factors $N_{2P}, N_{3P}, N_{2T}, N_{3T}$ read

$$N_{2P} = \sum_{i=1}^{M_{2P}} \exp\left(-\frac{\beta_{2Pi}^2}{2S_{2P}^2}\right), \quad N_{3P} = \sum_{j=1}^{M_{3P}} \exp\left(-\frac{\beta_{3Pj}^2}{2S_{3P}^2}\right), \tag{3}$$

$$N_{2T} = \sum_{l=1}^{M_{2T}} \exp\left(-\frac{\beta_{2Tl}^2}{2S_{2T}^2}\right), \quad N_{3T} = \sum_{n=1}^{M_{3T}} \exp\left(-\frac{\beta_{3Tn}^2}{2S_{3T}^2}\right). \tag{4}$$

In contrast to [23, 24], our present computer code allows to use different values for $M_{2P}, M_{3P}, M_{2T}, M_{3T}$. For each mode, the deformations are generated as follows:

$$\beta_{2Pi} = k_{2P} S_{2P} \left(\frac{2i}{M_{2P}-1} - 1\right), \qquad \beta_{3Pj} = k_{3P} S_{3P} \left(\frac{2j}{M_{3P}-1} - 1\right), \tag{5}$$

$$\beta_{2Tl} = k_{2T} S_{2T} \left(\frac{2l}{M_{2T}-1} - 1\right), \qquad \beta_{3Tn} = k_{3T} S_{3T} \left(\frac{2n}{M_{3T}-1} - 1\right). \tag{6}$$

Parameters $k_{2P}, k_{3P}, k_{2T}, k_{3T}$ are designed to account for the effective smearing of ZPO due to oscillations as the reagents approach each other. It seems plausible that ZPO does not show up for a "stiff" mode with small oscillations period. In contrast, a "soft" mode with large oscillations period will demonstrate ZPO in full power. From these considerations, we take the parameters $k_{2P}, k_{3P}, k_{2T}, k_{3T}$ as

$$k_{2P} = f \exp\left(-\frac{2\tau}{\tau_{2P}}\right), \quad k_{3P} = f \exp\left(-\frac{2\tau}{\tau_{3P}}\right), \tag{7}$$

$$k_{2T} = f \exp\left(-\frac{2\tau}{\tau_{2T}}\right), \quad k_{3T} = f \exp\left(-\frac{2\tau}{\tau_{3T}}\right). \tag{8}$$

The oscillations periods $\tau_{2P}, \tau_{3P}, \tau_{2T}, \tau_{3T}$ are related to the energies of first exited nuclear states $(E2+)_P, (E3-)_P, (E2+)_T, (E3-)_T$, respectively. Thus, we have two somewhat arbitrary variable parameters in the model, $1 < f < 3$ and $\tau \cong 1\ zs$. Parameter $\tau$ reflects to what extent ZPO survives when the reagents approach each other. Parameter $f$ allows to vary the range of accounted values of the deformations.

## III. NUCLEUS-NUCLEUS POTENTIAL

Since each of deformation parameters takes several values (see Eqs. (5), (6)), we restrict ourselves by the nose-to-nose geometry as in [5, 30-32]. In the DFM with the M3Y nucleon-nucleon (NN) forces for nose-to-nose geometry, the nucleus-nucleus potential $U$ reads

$$U(R, \hat{\beta}_P, \hat{\beta}_T) = U_C(R, \hat{\beta}_P, \hat{\beta}_T) + U_{nD}(R, \hat{\beta}_P, \hat{\beta}_T) + U_{nE}(R, \hat{\beta}_P, \hat{\beta}_T) \tag{9}$$

Here the Coulomb term $U_C$, the direct $U_{nD}$ and exchange $U_{nE}$ parts of SnnP $U_n$ depend upon the distance between the centers of mass of projectile and target nuclei $R$ and their deformations $\hat{\beta}_P = \{\beta_{2P}, \beta_{3P}\}$, $\hat{\beta}_T = \{\beta_{2T}, \beta_{3T}\}$. The three terms of Eq.(9) read [33]

$$U_C = \int d\vec{r}_P \int d\vec{r}_T \rho_{qP}(\vec{r}_P) v_C(s) \rho_{qT}(\vec{r}_T), \tag{10}$$

$$U_{nD} = \int d\vec{r}_P \int d\vec{r}_T \rho_{AP}(\vec{r}_P) v_D(s) \rho_{AT}(\vec{r}_T), \tag{11}$$

$$U_{nE} = \int d\vec{r}_P \int d\vec{r}_T \rho_{AP}(\vec{r}_P + \vec{s}) v_E(s) \rho_{AT}(\vec{r}_T - \vec{s}) \exp(i\vec{k}_{rel}\vec{s}/A_{red}). \tag{12}$$

Here $\rho_{AP}$ and $\rho_{AT}$ ($\rho_{qP}$ and $\rho_{qT}$) stand for the nucleon (charge) densities, $\vec{r}_P$ and $\vec{r}_T$ are the radius-vectors of the interacting points of the projectile and target nuclei,

$$\vec{s} = \vec{R} + \vec{r}_T - \vec{r}_P \tag{13}$$

The wave number $\vec{k}_{rel}$ is associated with the relative motion of colliding nuclei, the reduced mass number

$$A_{red} = \frac{A_P A_T}{A_P + A_T}. \tag{14}$$

Note that the structures of the DFM potentials using the M3Y- and Migdal NN forces are different. These potentials were compared in detail in [34].

In Eqs. (10)-(12), one neglects possible time-dependence of the densities (the so-called frozen densities approximation). This approximation seems to work reasonably well unless the density overlap of the colliding nuclei is about 1/3 of the saturation density 0.16 fm$^{-3}$ (see, e.g., [33]). Let us note that the density-dependent M3Y *NN*-forces are considered in the literature only for spherical colliding nuclei [35, 36]. It is not clear how to account for the density dependence when the nuclei are deformed. That is why in Eqs. (11), (12) we employ the bare (density-independent) *NN*-forces.

The direct part of the effective NN-interaction $v_D(s)$ consists of two Yukawa terms [25, 33, 37]:

$$v_D(s) = \sum_{i=1}^{2} G_{Di} \left[\exp(-s/r_{vi})\right]/(s/r_{vi}). \tag{15}$$

For the exchange part $v_E(s)$, one finds in the literature two options: an advanced and complicated one with the finite range (see e.g. [27, 35, 36]) and a simpler one with zero range [33]. The latter version reads:

$$v_E(s) = G_{E\delta}\,\delta(\vec{s}). \tag{16}$$

These are Eqs. (15), (16) which are employed in the present paper. The values of the coefficients are as follows: $G_{E\delta} = -592$ MeV fm$^3$, $G_{D1}$=11062 MeV, $G_{D2} = -2537.5$ MeV, $r_{v1}$=0.25 fm, $r_{v2}$=0.40 fm. Some details of calculations for the interaction energy of two deformed nuclei are discussed in Appendixes A1, A3.

For the nucleon and charge densities required for evaluating the DFM, we use the two-parameter Fermi formula (2pF-formula):

$$\rho_F(r) = \rho_{CF} \frac{1}{1 + \exp[(r - R_F)/a_F]}. \tag{17}$$

In Eq. (17), $R_F$ corresponds to the half central density radius, $a_F$ is the diffuseness, $\rho_{CF}$ is defined by the normalization condition. The 2pF-formula is often applied to approximate the experimental nuclear charge densities (see Ref. [38] and references therein).

To consider ZPO, we allow the spherical nuclei to become deformed. In this case, one should account for the density dependence upon the polar angle $\zeta$:

$$\rho_F(r, \zeta) = \rho_{CF} \frac{1}{1 + \exp\{[r - R_F h(\zeta)]/a_F\}} \tag{18}$$

where

$$h(\zeta) = \lambda^{-1} \left(1 + \beta_1 Y_{10}(\zeta) + \beta_2 Y_{20}(\zeta) + \beta_3 Y_{30}(\zeta)\right). \tag{19}$$

In Eq. (19), $\lambda$ is responsible for the volume conservation, $Y_{i0}$ are the spherical functions, $\beta_2$ and $\beta_3$ stand for the quadrupole and octupole deformation parameters, respectively, $\beta_1$ compensates the spurious center-of-mass shift resulting from non-zero $\beta_3$.

The half-density radii for proton $R_{Fp}$ and neutron $R_{Fn}$ densities as well as the corresponding diffusenesses $a_{Fp}$ and $a_{Fn}$ are taken from [39]. The half-density charge radius $R_{Fq}$ is taken to be approximately equal to $R_{Fp}$ whereas the charge diffuseness $a_{Fq}$ is calculated as [33]:

$$a_{Fq} = \sqrt{a_{FP}^2 + \frac{5}{7\pi^2}\left(0.76 - 0.11\frac{N}{Z}\right)}. \tag{20}$$

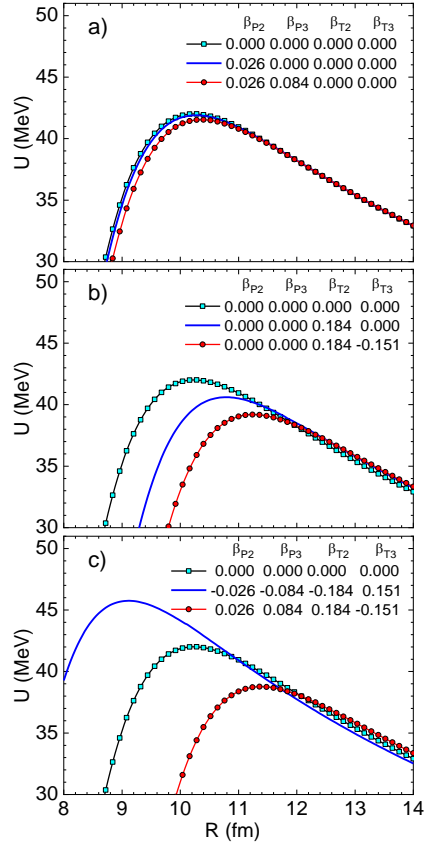

Fig. 1. Interaction energies $U(R)$ for several sets of deformations.
The values of deformation parameters are indicated in the figure and in Table 2.

Typical interaction energies $U(R)$ are presented for several sets of deformations in Fig. 1. In each panel, the nucleus-nucleus potential ignoring deformations is shown by blue lines with squares. In panel a) the curves with circles and without symbols denotes $U(R)$ calculated accounting for solely quadrupole and both quadrupole and octupole deformations of projectile nucleus, respectively. The values of $\beta_{P2}$, $\beta_{P3}$ are the maximum values calculated using Eq. (5). The effect of these deformations does not look significant that should be expected due to asymmetry of the reaction ($A_P \ll A_T$). Panel b) has analogous design, but for the target nucleus deformations. Here the barrier obtained with deformed target is significantly lower comparing to the case of both spherical nuclei. Note, that $\beta_{T3} < 0$ is chosen here to get the lowest possible barrier at fixed $\beta_{P2}$, $\beta_{P3}$, $\beta_{T2}$. Finally, in panel c) we show the cases with the lowest (curve with circles) and highest (curve without symbols) barriers. Although the corresponding absolute values of $\beta_{P2}$, $\beta_{P3}$, $\beta_{T2}$, $\beta_{T3}$ are rather moderate, the difference in the barrier heights is as large as 16%.

Table 2. The values of the parameters corresponding to the figures of the present work. $\beta_{P2m}$, $\beta_{P3m}$, $\beta_{T2m}$, $\beta_{T3m}$ denote the maximum values of the deformation parameters; "n/a" means "non applicable".

| Fig. | $f$ | $\tau$ (zs) | $K_R$ (zs/GeV) | $M_{2P}$ | $M_{3P}$ | $M_{2T}$ | $M_{3T}$ | $\beta_{2Pm}$ | $\beta_{3Pm}$ | $\beta_{2Tm}$ | $\beta_{3Tm}$ |
|---|---|---|---|---|---|---|---|---|---|---|---|
| 1 | 3.0 | 1.1 | n/a | n/a | n/a | n/a | n/a | 0.026 | 0.084 | 0.184 | 0.151 |
| 3 | 3.0 | 0.5 | 1 | varied | varied | varied | varied | 0.196 | 0.496 | 0.242 | 0.297 |
| 4 | 3.0 | varied | n/a | 3 | 3 | 3 | 3 | varied | varied | varied | varied |
| 5 | 3.0 | 1.1 | varied | 3 | 3 | 3 | 3 | 0.026 | 0.084 | 0.184 | 0.151 |
| 6 | 3.0 | 1.1 | 14 | 3 | 3 | 3 | 3 | 0.026 | 0.084 | 0.184 | 0.151 |
| 7 | 3.0 | varied | 14 | 3 | 3 | 3 | 3 | varied | varied | varied | varied |
| 8 | 3.0 | 1.1 | 14 | 3 | 3 | 3 | 3 | 0.026 | 0.084 | 0.184 | 0.151 |
| 9 | 2.0 | 1.1 | n/a | 3 | 3 | 3 | 3 | 0.026 | 0.084 | 0.184 | 0.151 |

# IV. OUTLINE OF THE TRAJECTORY MODEL ACCOUNTING FOR ZPO

The **Tra**jectory model accounting for dissipation, fluctuations, **B**arrier penetration, and **ZPO** for quadrupole and octupole modes is abbreviated as TRABZO. The logic of the model is similar but not identical to [40]; it is illustrated by Fig. 2.

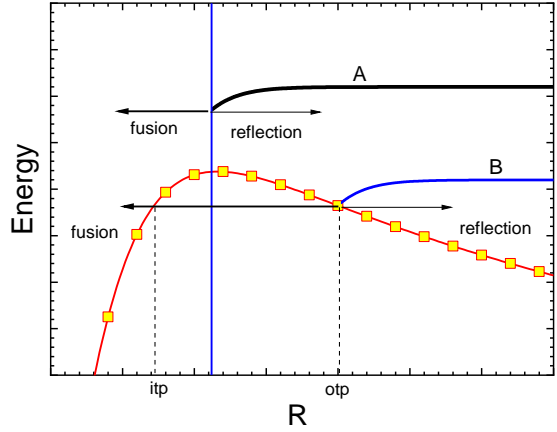

Fig. 2. Schematic illustration of TRABZO model. "itp" and "otp" denote the inner and outer turning points, respectively. For other details see text.

Within the framework of TRABZO, the relative motion of colliding nuclei is described as a motion of a fictious Brownian particle under the influence of conservative, dissipative, and fluctuating forces. Each trajectory is running until it either reaches the barrier radius having the collective energy $E_{coll}$ above the barrier (case A in Fig. 2) or just hits the slope of the barrier having $E_{coll} < B$ (case B). We calculate the probability to overcome the barrier using the parabolic barrier approximation for case A

$$T_J(\hat{\beta}_P, \hat{\beta}_T) = \left\{1 + \exp\left[\frac{2\pi(B_J(\hat{\beta}_P, \hat{\beta}_T) - E_{coll})}{\hbar\omega_{BJ}(\hat{\beta}_P, \hat{\beta}_T)}\right]\right\}^{-1} \quad (21)$$

and WKB-approximation in case B

$$T_J(\hat{\beta}_P, \hat{\beta}_T) = \left\{1 + \exp\left[\frac{2\Lambda_J(\hat{\beta}_P, \hat{\beta}_T)}{\hbar}\right]\right\}^{-1}. \quad (22)$$

Here $\Lambda_J(\hat{\beta}_P, \hat{\beta}_T)$ denotes the action calculated for the given deformations $\hat{\beta}_P, \hat{\beta}_T$ and angular momentum $J$ from the outer turning point ("otp") down to the inner one ("itp"). In Eq. (21), both the barrier energy $B$ and frequency $\omega_B$ are $J$ and $\beta_{2P}, \beta_{3P}, \beta_{2T}, \beta_{3T}$-dependent. Then we generate the uniform random number $\gamma$ between 0 and 1, and compare it with $T_J(\hat{\beta}_P, \hat{\beta}_T)$. Provided $T_J(\hat{\beta}_P, \hat{\beta}_T) < \gamma$ the trajectory is recorded as the scattered one, otherwise as the captured one. This algorithm results in $N_J(\hat{\beta}_P, \hat{\beta}_T)$ captured trajectories of $(2J + 1)N_0$ generated trajectories for each set $\{J, \hat{\beta}_P, \hat{\beta}_T\}$. The CS is calculated according to

$$\sigma_{th} = \sum_{i,j,k,l} \sigma_\beta \, \Pi_{P2i}\Pi_{P3j}\Pi_{2Tk}\Pi_{3Tl} \quad (23)$$

where

$$\sigma_\beta = \sigma(\hat{\beta}_P, \hat{\beta}_T) = \frac{\pi\hbar^2}{2m_R E_{c.m.}} \sum_J (2J+1) \frac{N_J(\hat{\beta}_P, \hat{\beta}_T)}{(2J+1)N_0}. \quad (24)$$

Typical values of $N_0$ range from 10 up to 100 for high and low collision energies, respectively. All details of the calculations and corresponding equations are collected in Appendix A2.

Our present approach resembles that of Ref. [41] and, as was already mentioned, of Ref. [40]. The differences of these three models are clarified in Table 1. Concerning our nose-to-nose approximation, we can argue that this geometry results in maximal CSs as demonstrated for instance in Fig. 3 of Ref. [41]. Of course, in the future development of the TRABZO, this approximation better to be avoided.

Table 1. Comparison between the preceding models [40], [41], and the present model TRABZO.

|  | [40] | [41] | **TRABZO (present work)** |
|---|---|---|---|
| **Deformations** | no | quadrupole, octupole | quadrupole, octupole |
| **Dynamical equations** | Langevin, dissipation and fluctuations | Hamilton, no dissipation | Langevin, dissipation and fluctuations |
| **Quantum fluctuations above the barrier** | no | no | yes |
| **Quantum tunneling** | yes | no | yes |
| **Orientations of colliding deformed nuclei** | no | different | nose-to-nose |
| **Strong nucleus-nucleus potential** | analytical profile [42] | double-folding with M3Y Reid NN-forces in linear approximation with respect to the deformation parameters | double-folding with M3Y Paris NN-forces exact with respect to the deformation parameters |

## V. RESULTS

All calculations in the present work are performed for reaction $^{16}$O+$^{92}$Zr, $S_{2P} = \beta_{2P}^0 = 0.349$, $S_{3P} = \beta_{3P}^0 = 0.729$, $S_{2T} = \beta_{2T}^0 = 0.101$, $S_{3T} = \beta_{3T}^0 = 0.174$ [29], but using $\tau$, $\tau_{2P}$ etc. (see Eqs. (7), (8)) significantly changes $\beta_{2P\max}$ etc.

Experimental CSs are taken from [43]. The only adjustable parameter in dynamical equations is the amplitude of the friction coefficient (friction strength) $K_R$. Increasing $K_R$ suppresses calculated CSs $\sigma_{th}$ whereas making $K_R$ extremely small (in practice, smaller than approximately 2 zs·GeV$^{-1}$) results in $\sigma_{th}$ equivalent to that coming from the barrier penetration model, $\sigma_{BPM}$, which represents the upper limit for the theoretical cross section. Of course, the latter can be calculated much easier using in Eq. (23) the values of $\sigma_\beta$ evaluated as follows:

$$\sigma_\beta = \sigma(\hat{\beta}_P, \hat{\beta}_T) = \frac{\pi\hbar^2}{2m_R E_{c.m.}} \sum_J (2J+1) T_J(\hat{\beta}_P, \hat{\beta}_T) \qquad (25)$$

instead of the dynamical calculations according to Eq. (24). The TRABZO calculations never have been performed before accounting for both quadrupole and octupole deformations of both colliding nuclei, therefore we prefer presenting all details in Figs. 3-7 (see also Figs. 9, 10 of Appendices). Everywhere, we use the ratio of the collision energy to the s-wave barrier energy for collision of spherical nuclei, $E_{c.m.}/B_{0sph}$, as an argument. The CSs calculated using Eq. (25) are denoted as $\sigma_{BPM}$, the CSs evaluated by means of the TRABZO are denoted as $\sigma_{TRA}$. The values of the variable parameters are collected in Table 2 in Appendix 1. To have a quantitative comparison with the data, we employ the ratios

$$\xi_{BPM} = \frac{\sigma_{BPM}}{\sigma_{exp}} \qquad (26)$$

and

$$\xi_{TRA} = \frac{\sigma_{TRA}}{\sigma_{exp}}. \qquad (27)$$

In Fig. 3, we present the ratios $\xi_{BPM}$ (panel a) and $\xi_{TRA}$ (panel b) calculated for low energies and five sets of deformations indicated in the figure. First, one sees that the BPM and TRABZO result in CSs which are hardly distinguishable. The reason is extremely weak friction $K_R = 1$ zs/GeV used for these calculations. This agrees with what was shown earlier (see Fig. 9 in Ref. [7]). Next, accounting for extra type of deformations results in increasing of the CSs at deep sub-barrier energies. This behavior agrees with Ref. [20]. Moreover, a careful examination of this figure demonstrates that deformations of $^{16}$O is significantly less important in comparison with the target deformations for this particular case.

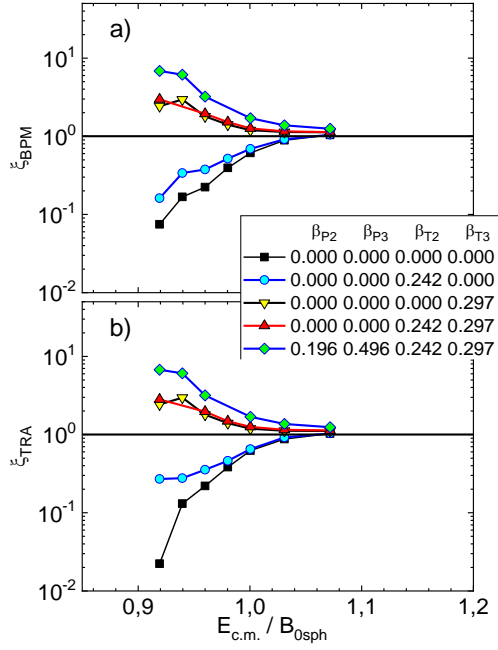

Fig. 3. The ratios $\xi_{BPM}$ (a) and $\xi_{TRA}$ (b) versus the ratio of collision energy to the s-wave spherical barrier energy.
These calculations are performed with $f = 3.0$, $\tau = 0.5$ zs, $K_R = 1.0$ zs/GeV.
The values of deformation parameters are indicated in the figure.

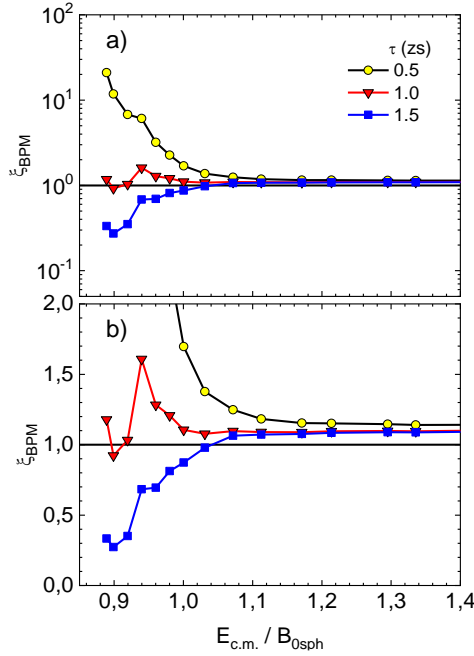

Fig. 4. The ratio $\xi_{BPM}$ in logarithmic (a) and linear (b) scales versus the ratio of collision energy to the s-wave spherical barrier energy for three values of $\tau$ which are indicated in the figure. $f = 3.0$.

Let us now study what is the impact of the adjustable parameter $\tau$ on the calculated cross sections. This study is illustrated by Fig. 4. Here $\xi_{BPM}$ evaluated for three values of $\tau$ are presented in logarithmic (panel a) and linear (panel b) scales. We see that for $E_{c.m.}/B_{0sph} > 1.1$ the value of $\tau$ is insignificant. For lower collision energies, this parameter influences the calculated CS strongly. It looks like $\tau \approx 1$ zs is promising for getting agreement between $\sigma_{TRA}$ and $\sigma_{exp}$ for the whole excitation function.

For the time being, we fix $\tau = 1.1$ zs and perform dynamical (TRABZO) calculations varying $K_R$ and calculating the value

$$\chi_\sigma^2 = \frac{1}{N_\sigma} \sum_{i=1}^{N_\sigma} \left[ \frac{\sigma_{TRAi} - \sigma_{expi}}{\Delta\sigma_{expi}} \right]^2. \tag{28}$$

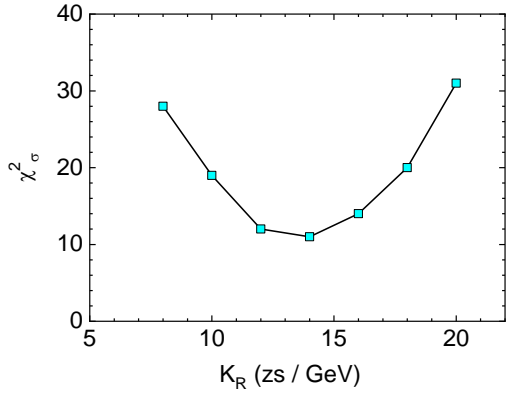

Fig. 5. $\chi^2_\sigma$ (see Eq. (28)) versus the friction strength coefficient. $f = 3.0$, $\tau = 1.1$ zs.

Results displayed in Fig. 5 indicate clear minimum $\chi^2_\sigma = 11$ at $K_R = 14$ zs GeV$^{-1}$. Note that from the systematics [11] one obtains for this reaction $K_R = 27$ zs GeV$^{-1}$. This is not a surprise: in that work, due to different NN-forces and nucleon densities, the barrier energy was $B_{0sph} = 41.6$ MeV whereas in the present work $B_{0sph} = 42.0$ MeV. Probably, for comparison we should make the TRABZO calculations with the ingredients of the DFM as in [11]. The problem is the difficulties in approximating the Hartree-Fock SKX densities used in [11] by the 2pF formula [44].

In Fig. 6, we compare $\sigma_{TRA}$ obtained using $K_R = 14$ zs GeV$^{-1}$ with $\sigma_{exp}$. The absolute values of the CSs are compared in linear and logarithmic scales in panels 6a and 6b, respectively. The ratios $\xi_{TRA}$ and $\xi_{BPM}$ shown in panel 6c enable one to get a more quantitative understanding. We believe the agreement is satisfactory for the whole excitation function. The role of friction at higher energies is clearly seen.

After getting good fit for $\sigma_{exp}$ with $K_R = 14$ zs GeV$^{-1}$ and $\tau = 1.1$ zs, we now slightly vary $\tau$ to be sure that this value indeed produces the best fit. Results presented in Fig. 7 demonstrate that this is the case. Also, this figure allows to estimate the impact of the value of $\tau$. Note, that here we present results of the computer-time consuming TRABZO calculations whereas in Fig. 4 results of fast BPM calculations are shown.

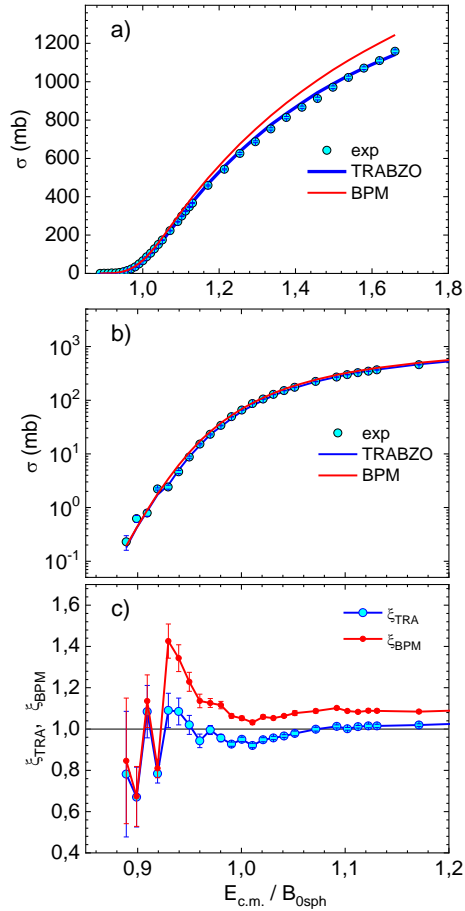

Fig. 6. Versus the ratio of collision energy to the s-wave spherical barrier energy are displayed: the CSs in linear (a) and logarithmic (b) scales as well as the ratios $\xi_{TRA}$, $\xi_{BPM}$ (c). $f = 3.0$, $\tau = 1.1$ zs, $K_R = 14$ zs/GeV.

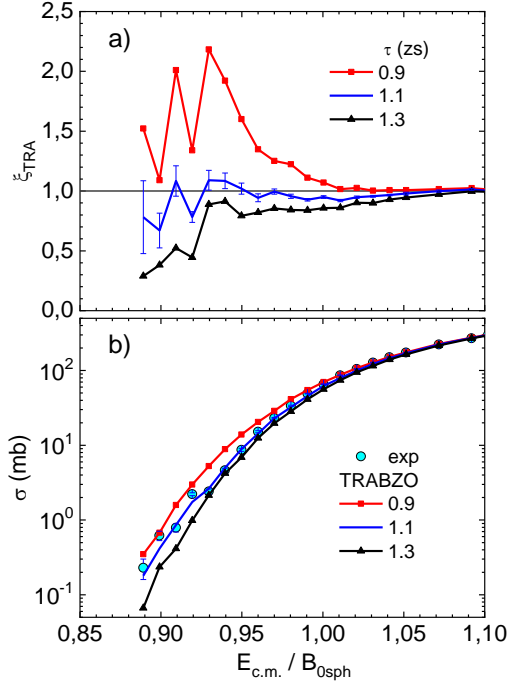

Fig. 7. The ratio $\sigma_{TRA}/\sigma_{exp}$ (a) and the CSs (b) versus the ratio of collision energy to the s-wave spherical barrier energy. Results of calculations are presented for three values of $\tau$ indicated in the figure.

In Refs. [3, 8], a method had been developed for extracting more information from the fusion excitation function: they calculated numerically the barrier distribution

$$D = \frac{d^2(E_{c.m.}\sigma)}{dE_{c.m.}^2}. \tag{29}$$

The experimental barrier distribution for our reaction, $^{16}O+^{92}Zr$, had been published in [43]. In Fig. 8, we compare the barrier distributions $D_{BPM}$ and $D_{TRA}$ resulting from our present calculations with the experimental ones $D_{exp}$. To simplify the numerical differentiation for this aim, we calculate the CSs with the constant energy step $\Delta E_{c.m.} = 0.5$ MeV. The differentiation is performed with the step $3\Delta E_{c.m.}$ for $E_{c.m.} < 47$ MeV and with the step $5\Delta E_{c.m.}$ otherwise as in [43]. We see that the barrier distributions obtained within BPM and TRABZO differ insignificantly although the CSs at high energies in Fig. 6 are noticeably different. Our calculated barrier distributions are in better agreement with the experimental ones than $D(E_{c.m.})$ calculated within the coupled-channels approach (see Fig. 3 in [43]).

The oscillations in the tail of $D_{TRA}$ are due to the stochastic nature of the TRABZO model. The values of $D_{BPM}$ behave smoothly with $E_{c.m.}$ as in calculations of Ref. [43].

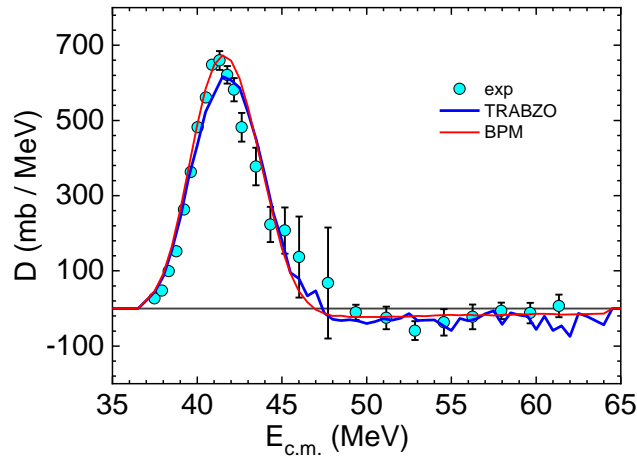

Fig. 8. The calculated (lines) and experimental (symbols, Ref. [43]) barrier distributions (see Eq. (29)) versus the collision energy.

# VI. CONCLUSIONS

Basing on the ideas of Refs. [20-22, 40], we have developed a model for evaluating the capture cross-sections (CSs) accounting for i) zero-point oscillations (ZPO) of the shape of colliding spherical nuclei and ii) dissipation and fluctuations of the radial coordinate $R$. Both quadrupole and octupole deformations of both colliding nuclei are accounted for. The nucleus-nucleus potential has been evaluated using the semi-microscopic double folding model for deformed nuclei in nose-to-nose geometry. The nucleon densities required for these calculations have been taken from Ref. [39]. For the effective nucleon-nucleon forces, the Paris M3Y forces with zero-range exchange part and original strength $-592$ MeV fm$^3$ from Ref. [25] have been used.

There are two options in the model for calculating the CSs. The simpler one, BPM, accounts only for ZPO, but does not involve any dynamics. The second one, TRABZO, is similar to the trajectory model with surface friction of Refs. [7, 11] except the final fate of a trajectory is decided by means of quantum transmission coefficients. Accordingly, we present two sorts of the calculated CSs, $\sigma_{BPM}$ and $\sigma_{TRA}$.

There are two fitting parameters in the model, $\tau$ (see Eqs. (7), (8)) and $K_R$ (see Eqs. (36), (37)). Parameter $\tau$ reflects to what extent ZPO survives when the reagents approach each other; it is included in the values of deformations for which the nucleus-nucleus interaction energy is calculated. Parameter $K_R$ is the friction strength for the radial motion. All calculations have been performed for $^{16}$O+$^{92}$Zr reaction. The calculated CSs and barrier distribution are compared with the precision data from [43] (typical experimental error is 1%).

Calculations show that the CSs are mostly sensitive to $\tau$ at low (sub-barrier) collision energies whereas the value of $K_R$ is important at the above barrier energies. Moreover, at sub-barrier energies $\sigma_{BPM} \approx \sigma_{TRA}$ whereas at above barrier energy the fusion process becomes suppressed by friction and $\sigma_{BPM} > \sigma_{TRA}$.

Good agreement between $\sigma_{exp}$ and $\sigma_{TRA}$ for the whole excitation function ($\chi^2_\sigma = 11$) has been obtained at $K_R = 14$ zs/GeV and $\tau = 1.1$ zs. The calculated barrier distribution is also in agreement with the experimental one. We hope the model can be upgraded and applied for fusion cross section in other reactions.

### Appendix A1. Number of multipoles

In practice, the integrals in Eqs. (10)-(12) are evaluated in two steps. First, one finds the multipole moments of the densities, erasing their dependence upon the polar angle $\zeta$. We consider up to 10 multipoles. The convergency of the resulting barrier energies with respect to the number of multipoles $L$ is illustrated by Fig. 9. Here calculations were performed with the values of the parameters indicated in Table 2.

The correspondence between the ordering number (horizontal axis) and the deformations is shown in Table 3. Altogether there are 81 sets of deformations. To make the effect of $L$ more visible, we plot the fractional difference

$$\xi_L = \frac{B_L}{B_{10}} - 1. \tag{30}$$

One could think that $L = 3$ is sufficient because only quadrupole and octupole deformations are considered (see e.g. [45]). However, according to Fig. 9 this is not the case. Even at $L = 5$ the fractional difference $\xi_L$ reaches 3%. Increasing the number of multipoles improves the situation, but only at $L > 7$ $\xi_L$ becomes smaller than 0.5%, the level we consider as acceptable. Note that in the previous studies [46, 47] where interaction of spherical projectile with deformed target was considered, up to $L = 10$ multipoles for expansion of the target densities was accounted for.

Table 3. The correspondence between $N$ (see Fig. 9) and the combination of deformations. $\beta_{P2m}$, $\beta_{P3m}$, $\beta_{T2m}$, $\beta_{T3m}$ denote the maximum values of the deformation parameters.

| N | $\beta_{2P}$ | $\beta_{3P}$ | $\beta_{2T}$ | $\beta_{3T}$ | N | $\beta_{2P}$ | $\beta_{3P}$ | $\beta_{2T}$ | $\beta_{3T}$ |
|---|---|---|---|---|---|---|---|---|---|
| 1,2,3 | $-\beta_{2Pm}$ | $-\beta_{3Pm}$ | $-\beta_{2Tm}$ | $-\beta_{3Tm}, 0, +\beta_{3Tm}$ | 43,44,45 | 0 | 0 | $+\beta_{2Tm}$ | $-\beta_{3Tm}, 0, +\beta_{3Tm}$ |
| 4,5,6 | $-\beta_{2Pm}$ | $-\beta_{3Pm}$ | 0 | $-\beta_{3Tm}, 0, +\beta_{3Tm}$ | 46,47,48 | 0 | $+\beta_{3Pm}$ | $-\beta_{2Tm}$ | $-\beta_{3Tm}, 0, +\beta_{3Tm}$ |
| 7,8,9 | $-\beta_{2Pm}$ | $-\beta_{3Pm}$ | $+\beta_{2Tm}$ | $-\beta_{3Tm}, 0, +\beta_{3Tm}$ | 49,50,51 | 0 | $+\beta_{3Pm}$ | 0 | $-\beta_{3Tm}, 0, +\beta_{3Tm}$ |
| 10,11,12 | $-\beta_{2Pm}$ | 0 | $-\beta_{2Tm}$ | $-\beta_{3Tm}, 0, +\beta_{3Tm}$ | 52,53,54 | 0 | $+\beta_{3Pm}$ | $+\beta_{2Tm}$ | $-\beta_{3Tm}, 0, +\beta_{3Tm}$ |
| 13,14,15 | $-\beta_{2Pm}$ | 0 | 0 | $-\beta_{3Tm}, 0, +\beta_{3Tm}$ | 55,56,57 | $+\beta_{2Pm}$ | $-\beta_{3Pm}$ | $-\beta_{2Tm}$ | $-\beta_{3Tm}, 0, +\beta_{3Tm}$ |
| 16,17,18 | $-\beta_{2Pm}$ | 0 | $+\beta_{2Tm}$ | $-\beta_{3Tm}, 0, +\beta_{3Tm}$ | 58,59,60 | $+\beta_{2Pm}$ | $-\beta_{3Pm}$ | 0 | $-\beta_{3Tm}, 0, +\beta_{3Tm}$ |
| 19,20,21 | $-\beta_{2Pm}$ | $+\beta_{3Pm}$ | $-\beta_{2Tm}$ | $-\beta_{3Tm}, 0, +\beta_{3Tm}$ | 61,62,63 | $+\beta_{2Pm}$ | $-\beta_{3Pm}$ | $+\beta_{2Tm}$ | $-\beta_{3Tm}, 0, +\beta_{3Tm}$ |
| 22,23,24 | $-\beta_{2Pm}$ | $+\beta_{3Pm}$ | 0 | $-\beta_{3Tm}, 0, +\beta_{3Tm}$ | 64,65,66 | $+\beta_{2Pm}$ | 0 | $-\beta_{2Tm}$ | $-\beta_{3Tm}, 0, +\beta_{3Tm}$ |
| 25,26,27 | $-\beta_{2Pm}$ | $+\beta_{3Pm}$ | $+\beta_{2Tm}$ | $-\beta_{3Tm}, 0, +\beta_{3Tm}$ | 67,68,69 | $+\beta_{2Pm}$ | 0 | 0 | $-\beta_{3Tm}, 0, +\beta_{3Tm}$ |
| 28,29,30 | 0 | $-\beta_{3Pm}$ | $-\beta_{2Tm}$ | $-\beta_{3Tm}, 0, +\beta_{3Tm}$ | 70,71,72 | $+\beta_{2Pm}$ | 0 | $+\beta_{2Tm}$ | $-\beta_{3Tm}, 0, +\beta_{3Tm}$ |
| 31,32,33 | 0 | $-\beta_{3Pm}$ | 0 | $-\beta_{3Tm}, 0, +\beta_{3Tm}$ | 73,74,75 | $+\beta_{2Pm}$ | $+\beta_{3Pm}$ | $-\beta_{2Tm}$ | $-\beta_{3Tm}, 0, +\beta_{3Tm}$ |
| 34,35,36 | 0 | $-\beta_{3Pm}$ | $+\beta_{2Tm}$ | $-\beta_{3Tm}, 0, +\beta_{3Tm}$ | 76,77,78 | $+\beta_{2Pm}$ | $+\beta_{3Pm}$ | 0 | $-\beta_{3Tm}, 0, +\beta_{3Tm}$ |
| 37,38,39 | 0 | 0 | $-\beta_{2Tm}$ | $-\beta_{3Tm}, 0, +\beta_{3Tm}$ | 79,80,81 | $+\beta_{2Pm}$ | $+\beta_{3Pm}$ | $+\beta_{2Tm}$ | $-\beta_{3Tm}, 0, +\beta_{3Tm}$ |
| 40,41,42 | 0 | 0 | 0 | $-\beta_{3Tm}, 0, +\beta_{3Tm}$ | | | | | |

In the real DFM calculations, after finding the multipole components of the densities, for the second step, one performs their Fourier transform. Finally, the multipole components of the nucleus-nucleus interaction are evaluated. In these calculations, one needs spherical Bessel functions $j_l(x)$ up to $2L$-order and large values of argument ($x \leq 200$). The table of these functions with the step $\Delta x = 0.05$ was obtained using Eqs. (10.1.8) and (10.1.2) from Ref. [48] as was proposed in [49]. Other details of the calculations and collections of formulas can be found in [45, 50].

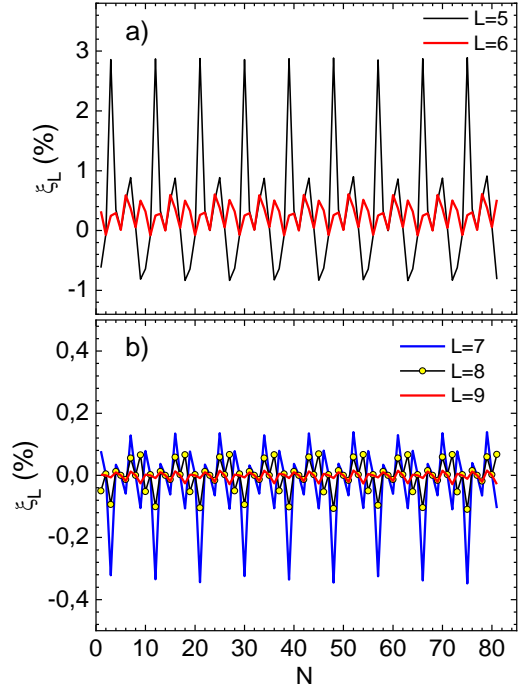

Fig. 9. Fractional difference between the barrier energy calculated at the given $L$ and $L = 10$ (see Eq. (30)). The number $N$ corresponding to the abscissa characterizes the deformations set (see Table 3). $f = 2$, $\tau = 1.1$ zs.

### Appendix A2. Modeling of individual trajectory within fluctuation-dissipation model

The physical picture of modeling individual trajectory is essentially the one of Brownian motion in external field. The fictitious particle with the reduced mass wanders experiencing the action of the conservative, dissipative, and random (fluctuating) forces. This motion is described by the dimensionless coordinate $q$ which is proportional to the distance between the centers of mass of the projectile and target nuclei $R$:

$$q = R \left[ \lambda_P^{-1}(1 + \beta_{1P}Y_{10}(0) + \beta_{2P}Y_{20}(0) + \beta_{3P}Y_{30}(0)) + \lambda_T^{-1}(1 + \beta_{1T}Y_{10}(0) + \beta_{2T}Y_{20}(0) + \beta_{3T}Y_{30}(0)) \right]^{-1}. \quad (31)$$

The memory effects are neglected therefore we use the stochastic Langevin-type equations with the white noise and instant dissipation:

$$dp = (F_U + F_{\text{cen}} + F_{fr})dt + \sqrt{2D_\theta}\, dW, \quad (32)$$

$$dq = p\, dt/m_q, \quad (33)$$

$$F_U = -\frac{dU}{dq}, \quad (34)$$

$$F_{\text{cen}} = \frac{\hbar^2 J^2}{m_q q^3}, \quad (35)$$

$$F_{fr} = -\frac{p}{m_q} K_R \left[\frac{dU_n}{dq}\right]^2, \quad (36)$$

$$D_\theta = \theta K_R \left[\frac{dU_n}{dq}\right]^2. \quad (37)$$

Here $p$ denotes the linear momentum corresponding to the radial motion; $F_U$, $F_{\text{cen}}$, and $F_{fr}$ are the conservative, centrifugal, and frictional forces, respectively. The latter is related to the nucleus-nucleus strong interaction potential $U_n(q)$ via the surface friction expression [5, 42]. $U(q)$ is the total nucleus-nucleus interaction energy consisting of the Coulomb $U_C(q)$ and nuclear $U_n(q)$ parts; $J\hbar$ is the projection of the orbital angular momentum onto the axis perpendicular to the reaction plane; $m_q$ is the inertia parameter;

$K_R$ denotes the dissipation strength coefficient; $D_\theta$ stands for the diffusion coefficient which is proportional to the temperature $\theta$. The random force is proportional to the increment of the Wiener process $dW$, the latter possesses zero average and variance equal to $dt$. Equations (32), (33) are solved numerically using the Runge-Kutta method (see details in [51, 52]).

### Appendix A3. Constructing the SnnP at large center-to-center distances

Computing the frictional force $F_{fr}$ (Eq. (36)) requires the smooth strong nucleus-nucleus potential (SnnP). However, at large distances between the centers of masses (strictly speaking, between nuclear surfaces), the calculations within the framework of DFM lose the accuracy as was discussed in [7]. Therefore, we approximate the calculated SnnP by the Gross-Kalinowski (GK) profile (see Fig. 5 in Ref. [7])

$$U_{nGK} = \ln\left[1 + \exp\left(-\frac{\Delta R}{a_{GK}}\right)\right](A_{0GK} + A_{1GK}\Delta R + A_{2GK}\Delta R^2), \tag{38}$$

$$\Delta R = R - r_{GK}(A_P^{1/3} + A_T^{1/3}). \tag{39}$$

In the present work, the coefficients of this approximation for each set of deformations were optimized in the interval from $R_m = R_B + 1$ fm up to $U_n = -0.01$ MeV. The GK profile with such found coefficients was then used for all $R > R_m$ whereas at smaller values of $R$ the DFM SnnP was used. The quality of this routine is illustrated by Fig. 10.

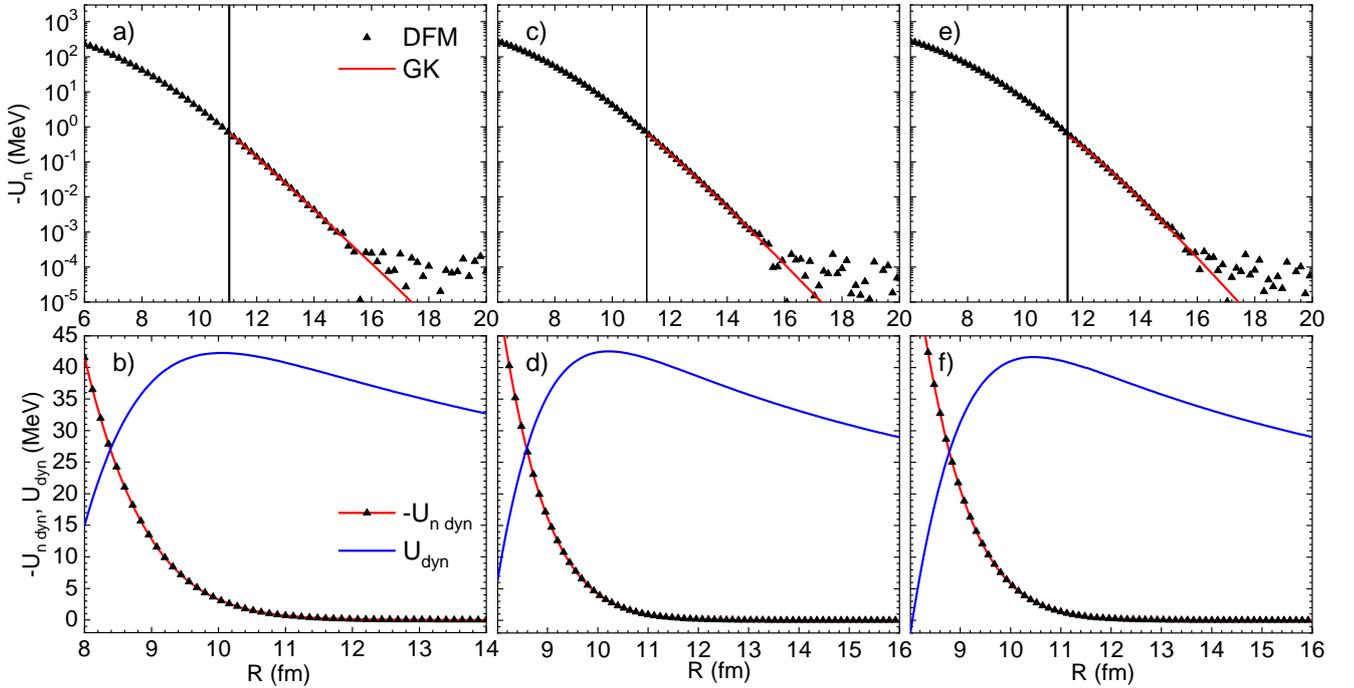

Fig. 10. Upper row: the absolute value of SnnP calculated using the DFM (symbols) and the GK approximations (lines) in the logarithmic scale. Vertical lines correspond to $R_m$ (see text). Lower row: the absolute value of SnnP and total interaction energy $U_{dyn}$ applied for the dynamical calculations in Eqs. (32)-(37). The deformations and barrier energies are as follows:
a) and b) $\beta_{P2} = -0.026$, $\beta_{P3} = -0.084$, $\beta_{T2} = -0.184$, $\beta_{T3} = -0.151$, $B_0 = 42.28$ MeV;
c) and d) $\beta_{P2} = -0.026$, $\beta_{P3} = -0.084$, $\beta_{T2} = +0.184$, $\beta_{T3} = +0.151$, $B_0 = 42.55$ MeV;
e) and f) $\beta_{P2} = +0.026$, $\beta_{P3} = +0.084$, $\beta_{T2} = +0.184$, $\beta_{T3} = +0.151$, $B_0 = 41.65$ MeV.

In panels a), c), e), the values of $-U_{nDF}(R)$ are shown by symbols whereas lines are used for the approximating GK-profiles. Vertical lines correspond to the values of $R_m$. We see that at relatively large $R$-values (in other words at small values of $-U_{nDF}$) the DFM-procedure loses its accuracy. The value

$$\chi_U^2 = \frac{1}{N_U}\sum_{i=1}^{N_U}\left[\frac{U_{GK}(R_i) - U_{nDF}(R_i)}{U_{GK}(R_i) + U_{nDF}(R_i)}\right]^2 \tag{40}$$

characterize quantitatively the agreement between $U_{nDF}(R)$ and $U_{nGK}(R)$. Here $N_U$ is the number of points employed for finding the coefficients of GK-profile. Typical values of $\chi_U^2$ is $10^{-4}$ or less.

In panels b), d), e), the values of $-U_{ndyn}(R)$ and $U_{dyn}(R)$ used for calculating the capture cross sections are shown. This SnnP (red lines with triangles) is constructed as described above. One sees extremely smooth behavior of $U_{ndyn}(R)$ which provides the accurate total nucleus-nucleus interaction energies $U_{dyn}(R)$ (blue lines without symbols). Note that we select the most "suspicious" sets where each reagent possesses both type of deformations. The problem with oscillations observed in panels a), c), e) for $U_{nDF}(R)$ appears for the Coulomb interaction $U_{CDF}(R)$ at much larger distance therefore at these distances $U_{CDF}(R)$ is changed to the monopole-monopole Coulomb interaction.

## Acknowledgements

This work was supported by the Foundation for the Advancement of Theoretical Physics and Mathematics "BASIS". The author is indebted to Prof. I. I. Gontchar for valuable discussions.